\documentclass[final,3p,times,twocolumn]{elsarticle}
\usepackage{xcolor}
\usepackage{url}
\usepackage{amssymb}
\usepackage{adjustbox}
\usepackage{subfigure}
\journal{Journal of High Energy Astrophysics}
\usepackage{amsmath}
\begin{document}

\begin{frontmatter}

%% Title, authors and addresses

%% use the tnoteref command within \title for footnotes;
%% use the tnotetext command for theassociated footnote;
%% use the fnref command within \author or \affiliation for footnotes;
%% use the fntext command for theassociated footnote;
%% use the corref command within \author for corresponding author footnotes;
%% use the cortext command for theassociated footnote;
%% use the ead command for the email address,
%% and the form \ead[url] for the home page:
%% \title{Title\tnoteref{label1}}
%% \tnotetext[label1]{}
%% \author{Name\corref{cor1}\fnref{label2}}
%% \ead{email address}
%% \ead[url]{home page}
%% \fntext[label2]{}
%% \cortext[cor1]{}
%% \affiliation{organization={},
%%             addressline={},
%%             city={},
%%             postcode={},
%%             state={},
%%             country={}}
%% \fntext[label3]{}

\title{A White Dwarf Binary Candidate Discovered by LAMOST Using Dynamical Method }

%\author[inst1]{Author One}
\author[a]{Haifan Zhu}
\author[a]{Wei Wang}
\ead{wangwei2017@whu.edu.cn}
\author[b,d]{Xue Li}
\author[c,d]{Jia-jia Li}
\author[a]{Pengfu Tian}
\address[a]{Department of Astronomy, School of Physics and Technology, Wuhan University, Wuhan 430072, China}
\address[b]{Key Laboratory of Optical Astronomy, National Astronomical Observatories, Chinese Academy of Sciences, Beijing 100101, China}
\address[c]{Yunnan Observatories, Chinese Academy of Sciences, Kunming, 650216, China}
\address[d]{College of Astronomy and Space Sciences, University of Chinese Academy of Sciences, Beijing 100049, China}

\begin{abstract}
We present the discovery of a binary system containing a white dwarf candidate using data from the LAMOST. Our analysis of the radial velocity data allowed us to determine an orbital period of approximately 0.953 days and a mass function of 0.129 $M_\odot$. Through spectral energy distribution (SED) fitting, we obtained the stellar parameters of the visible star. By combining these results with the mass function, we established a relationship between the mass of the invisible star and the system's inclination angle, along with the Roche lobe radius. We find that the mass of the invisible star is below the Chandrasekhar limit when the inclination angle exceeds $35^\circ$. Given that systems with large variations in radial velocity typically have high inclination angles, we classify the invisible star as a white dwarf candidate. The Roche lobe radius exceeds the physical radius of the visible star, indicating that no mass transfer occurs, which results in a weak ellipsoidal modulation effect. Additionally, we obtained light curves from the TESS, ASAS-SN, and CRTS surveys. The light curves also exhibit a periodicity of approximately 0.95 days, with ellipsoidal modulation only in the 2019 TESS observations. Coupled with the strong $\rm H_{\alpha}$ emission line observed in the LAMOST MRS spectrum, we infer that the surface of the visible star contains significant hot spots. This obscures the system's inherently weak ellipsoidal modulation, resulting in a manifestation of rotational variables. Furthermore, an analysis of the dynamical characteristics of this system indicates that it has a high inclination angle  ($>60$ degrees) and its orbital properties are consistent with those of typical thin disk stars, supporting the hypothesis that the invisible object is a white dwarf.

\end{abstract}

\begin{keyword}
%% keywords here, in the form: keyword \sep keyword
Radial velocity \sep binary stars \sep white dwarf
%% PACS codes here, in the form: \PACS code \sep code

%% MSC codes here, in the form: \MSC code \sep code
%% or \MSC[2008] code \sep code (2000 is the default)

\end{keyword}

\end{frontmatter}

%% Add \usepackage{lineno} before \begin{document} and uncomment 
%% following line to enable line numbers
%% \linenumbers

%% main text
%%

%% Use \section commands to start a section
\section{Introduction}
\label{sec1}
%% Labels are used to cross-reference an item using \ref command.

At the end of their evolutionary life cycles, stars undergo processes that lead to the formation of compact objects, including white dwarfs (WDs), neutron stars (NSs), and black holes (BHs). The evolutionary pathways that determine these outcomes are primarily dictated by initial stellar parameters such as mass, metallicity, and rotational velocity. The identification and detailed characterization of these compact remnants are crucial for advancing our understanding of stellar evolution. These objects offer unique insights into extreme astrophysical environments characterized by strong gravitational fields and high-density matter, while also providing valuable information on binary interactions and the broader interstellar medium. 
In recent years, significant attention has been directed toward the search for non-interacting binaries that include a compact object \citep{mullally2009twins,casewell2018first,jayasinghe2023search}. More than half of the stars in the galaxy are found in binary systems \citep{duchene2013stellar,whitworth2015majority}, many of which contain a significant population of compact objects\citep{remillard2006,rebassa2012post,chen2023binary}. 

Radial velocity (RV) monitoring has demonstrated its effectiveness in uncovering invisible massive compact stars in binary systems. 
The Large Sky Area Multi-Object Fiber Spectroscopic Telescope (LAMOST), also known as the GuoShouJing Telescope, is a powerful reflecting Schmidt telescope with a four-meter effective aperture and a wide field of view. With the capability to utilize approximately 4,000 fibers, LAMOST contributes over 1,000,000 spectra annually, making it one of the most advanced optical spectroscopic survey telescopes in operation\citep{cui2012large,zhao2012lamost}. The observation strategy of LAMOST , which involves acquiring multiple spectra for individual targets, offers an excellent opportunity to identify binary systems containing compact objects through RV monitoring methods. This approach has facilitated the discovery of a significant number of black hole binaries \citep{gu2019method,liu2019wide,wang2024potential}. This method can also be used to search for neutron star binaries\citep{swihart2021discovery,yi2022dynamically} and white dwarf binaries \citep{el2021lamost,li2022,qi2023searching,zheng2023nearest,rowan2024hidden}. 

In some cases, the non-interacting black hole binaries that have been identified before do not actually consist of black holes. The study analyzes two binary systems with giant stars, V723 Mon and 2M04123153+6738486, both of which were initially thought to contain mass-gap black holes\citep{jayasinghe2021unicorn,jayasinghe2022giraffe}. However, spectral disentangling reveals that these systems host luminous companions with star-like spectra. Joint modeling indicates that the primary components are luminous, cool giants with low masses (approximately 0.4 $M_{\odot}$) that likely fill their Roche lobes, while the secondary components are slightly warmer subgiants. The characteristics of both systems are consistent with binary evolution models, suggesting they represent a rarely observed phase before the formation of wide-orbit helium white dwarfs and ultimately compact binaries containing two helium white dwarfs\cite{el2022unicorns}.

The search for massive white dwarfs can also encounter similar issues. Measuring their mass distribution is crucial for studying asymptotic giant branch pulsations and the chemical evolution of galaxies \cite{camisassa2019evolution}. The mass distribution of white dwarfs displays a dominant peak near $ M_{\text{WD}} \sim 0.6 \, M_\odot$, accompanied by a smaller peak at the tail of the distribution, approximately $ M_{\text{WD}} \sim 0.82 \, M_\odot $ \cite{camisassa2019evolution, kleinman2012sdss}.
Another study revisits the properties and implications of the white dwarf mass distribution \cite{tremblay2016field}. Using the most complete and precise atmospheric parameters from the 20 pc volume-complete survey and the Sloan Digital Sky Survey magnitude-limited sample, the authors modeled the observed mass distributions with Monte Carlo simulations. Despite various biases and uncertainties, they found that standard assumptions resulted in predicted mean masses that were in good agreement with the observed values. However, their simulations over-predicted the number of massive white dwarfs (\(M > 0.75 \, M_\odot\)) by 40–50\%, and they found no evidence for a population of double white dwarf mergers \cite{tremblay2016field}. 

The binary systems containing compact objects are important for binary evolution, while the sample identified by the dynamical method is relatively small. Thus, searching for these binary systems of unique properties and behaviors in the LAMOST spectra would be promising for revealing more compact objects. 
In this paper, we report a white dwarf binary system with K-dwarf companions, identified through radial velocity observations derived from LAMOST data. 
This paper is organized as follows: we present the data observation and selection in Section~\ref{sec2}; Section~\ref{sec3} presents the results of the binary star analysis, including the radial velocity fitting results for the visible star, the fitting of fundamental parameters, and the calculated parameters for the binary system; finally, we give the discussion and conclusions in Sections~\ref{sec4} and ~\ref{sec5}.

\section{Data Observation and Selection}
\label{sec2}
We utilized the latest LAMOST Data Release 11 (DR11) V1.0\footnote{\url{https://www.lamost.org/dr11/v1.0/}}. DR11 offers multiple spectral observations for individual targets in both the Low Resolution Survey (LRS) and the Medium Resolution Survey (MRS). The Medium Resolution Spectra has a resolution of 7500 at 5163 Å (blue band) and 6593 Å (red band), enabling precise measurement of radial velocities and enhanced identification of spectral components. This dataset includes 10,432,303 spectra from the MRS Time-Domain Survey. 

We filtered the data from the complete DR11 star catalog 
( namely {\tt dr11\_v1.0\_MRS\_catalogue}) by first removing observations flagged with fiber issues. We then applied the following selection criteria: observations with more than 20 instances, ensuring that the radial velocity measurements encompass most of the orbital phases, clear radial velocity variations greater than 20 $\rm km~s^{-1}$, and exclusion of double-lined spectroscopic binaries. We first used the Lomb–Scargle method \citep{lomb1976least,scargle1981studies} to search for periodicity in the selected data with radial velocity variations, identifying sources with a false alarm probability greater than 0.005. Subsequently, we derived the orbital solution for these target sources by fitting the radial velocity data using the Joker code\footnote{\url{https://github.com/adrn/thejoker}} \citep{price2017joker}. We examined the results after folding the periods individually and retained sources with clear periodic signals. 

Based on the parameters fitted by the Joker code, we can calculate the mass function of the binary star: 
\begin{equation}
    f(M) = \frac{M_2 \sin^3 i}{\left(1+q \right)^2} = \frac{P K^3(1 - e^2)^{3/2}}{2 \pi G} ,
\end{equation}
where \( q \) represents the mass ratio between the two stars, defined as \( M_1/M_2 \), where \( M_1 \) is the mass of the visible star and \( M_2 \) is the mass of the invisible one. The inclination angle of the system is denoted by \( i \), \( P \) is the orbital period, \( e \) is the eccentricity, and \( K \) is the semi-amplitude of the radial velocity. 

We filtered the final sample by selecting sources with a mass function greater than 0.1. In our sample, some sources have already been studied in details in previous work, therefore, we excluded these sources at first \citep{li2022binaries,qi2023searching}. Finally, we performed a cross-match with the Double-lined Spectroscopic Binary catalog \citep{zhang2022spectroscopic} to exclude the double-lined spectroscopic binaries. The remaining sources were then individually inspected to ensure that they are single-lined spectroscopic binaries. During this process, white dwarf binary candidate with strong H$_\alpha$ emission lines, named as LAMOST J230854.08+355132.4 (hereafter J2308). The observational data we selected are listed in Table~\ref{tab1}. 
\begin{table*}
%\bc
\centering
%\begin{minipage}[]{100mm}
\caption{ LAMOST MRS observations for J2308.\label{tab1} LMJD  is obtained through the conversion of Local Modified Julian Minute. RVB, RVR, $\sigma_{RVB}$, and $\sigma_{RVR}$ represent the radial velocities obtained from the blue arm and the red arm, respectively, along with their corresponding errors. All values were obtained from the MRS catalog.}
  \renewcommand{\arraystretch}{0.85}
 \setlength{\tabcolsep}{4.5pt}
 \begin{tabular}{cccccc|cccccc}
 \hline\noalign{\smallskip}
 LMJD &  RVB & $\sigma_{RVB}$ & RVR &$\sigma_{RVR}$ & {\it SNR} &  lMJD &  RVB & $\sigma_{RVB}$ & RVR &$\sigma_{RVR}$ & {\it SNR}\\
 (day) & ($\rm km~s^{-1}$) & ($\rm km~s^{-1}$) & ($\rm km~s^{-1}$) & ($\rm km~s^{-1}$) && (day) & ($\rm km~s^{-1}$) & ($\rm km~s^{-1}$) & ($\rm km~s^{-1}$) & ($\rm km~s^{-1}$)  & \\
 \hline\noalign{\smallskip}
59542.784 & -70.19  & 3.07 & -72.77  & 3.28 & 20.64 & 59868.911 & -117.34 & 2.22 & -116.25 & 2.35 & 48.48 \\
59542.799 & -76.24  & 3.38 & -81.71  & 3.04 & 23.11 & 59888.874 & -93.71  & 2.86 & -96.59  & 2.88 & 21.56 \\
59542.830 & -102.15 & 2.99 & -98.09  & 3.21 & 20.45 & 59888.859 & -86.20  & 2.70 & -87.36  & 2.62 & 27.91 \\
59186.793 & 75.63   & 3.19 & 77.88   & 2.92 & 25.07 & 59888.844 & -80.86  & 2.51 & -83.67  & 2.59 & 28.80 \\
59186.826 & 93.69   & 3.64 & 86.61   & 2.97 & 22.01 & 59888.829 & -67.82  & 2.63 & -71.40  & 2.49 & 30.34 \\
59126.902 & 82.49   & 2.58 & 82.38   & 2.25 & 35.47 & 59888.814 & -63.20  & 2.66 & -65.06  & 2.44 & 31.24 \\
59126.919 & 72.98   & 2.64 & 73.74   & 2.55 & 31.39 & 59888.799 & -49.28  & 2.62 & -51.92  & 2.63 & 36.39 \\
58801.838 & 67.49   & 2.92 & 74.58   & 2.54 & 36.15 & 59888.783 & -44.01  & 2.50 & -43.11  & 2.42 & 38.85 \\
58801.772 & 83.24   & 3.43 & 91.70   & 2.71 & 25.57 & 59186.777 & 63.88   & 2.81 & 74.79   & 2.88 & 26.38 \\
58801.805 & 87.87   & 2.87 & 79.45   & 2.56 & 31.70 & 59186.809 & 87.83   & 3.04 & 82.07   & 3.23 & 22.71 \\
58801.821 & 77.93   & 2.44 & 75.37   & 2.70 & 36.39 & 59870.932 & -117.92 & 2.44 & -116.76 & 2.71 & 38.65 \\
59176.790 & -110.67 & 2.97 & -111.85 & 2.97 & 23.01 & 59870.917 & -120.97 & 2.39 & -120.62 & 2.74 & 41.79 \\
59176.774 & -108.26 & 2.74 & -101.71 & 2.91 & 25.78 & 59870.901 & -120.65 & 2.45 & -119.69 & 2.46 & 45.13 \\
59868.957 & -124.28 & 2.49 & -126.63 & 2.66 & 38.07 & 59870.886 & -124.87 & 2.37 & -128.51 & 2.48 & 45.21 \\
59868.942 & -126.30 & 2.55 & -129.10 & 2.73 & 40.80 & 59868.896 & -110.30 & 2.31 & -112.93 & 2.39 & 45.54 \\
59868.926 & -123.20 & 2.38 & -125.85 & 2.53 & 42.75 &  & & & & & \\
 \noalign{\smallskip}\hline
  \end{tabular}
 % \begin{tablenotes}
  %      \item[a] lMJD  is obtained through the conversion of Local Modified %Julian Minute.
   %     \item[b] QPO represents the centroid frequency of QPOs
   % \end{tablenotes}
  \end{table*}
\section{Result of Binary Analysis}
\label{sec3}
\subsection{Radial Velocity Curves}
We used the radial velocities from the blue arm of the LAMOST MRS for our analysis. We performed a Keplerian fit using the custom Markov Chain Monte Carlo (MCMC) sampler, the Joker, with a total length of 30,000 steps. The error statistics were calculated using the 16th percentile and 84th percentile of the sample distribution, representing the lower and upper limits of the confidence interval, respectively. This corresponds to approximately a 68\% confidence level.  

 \begin{figure}

    \includegraphics[width=\columnwidth]{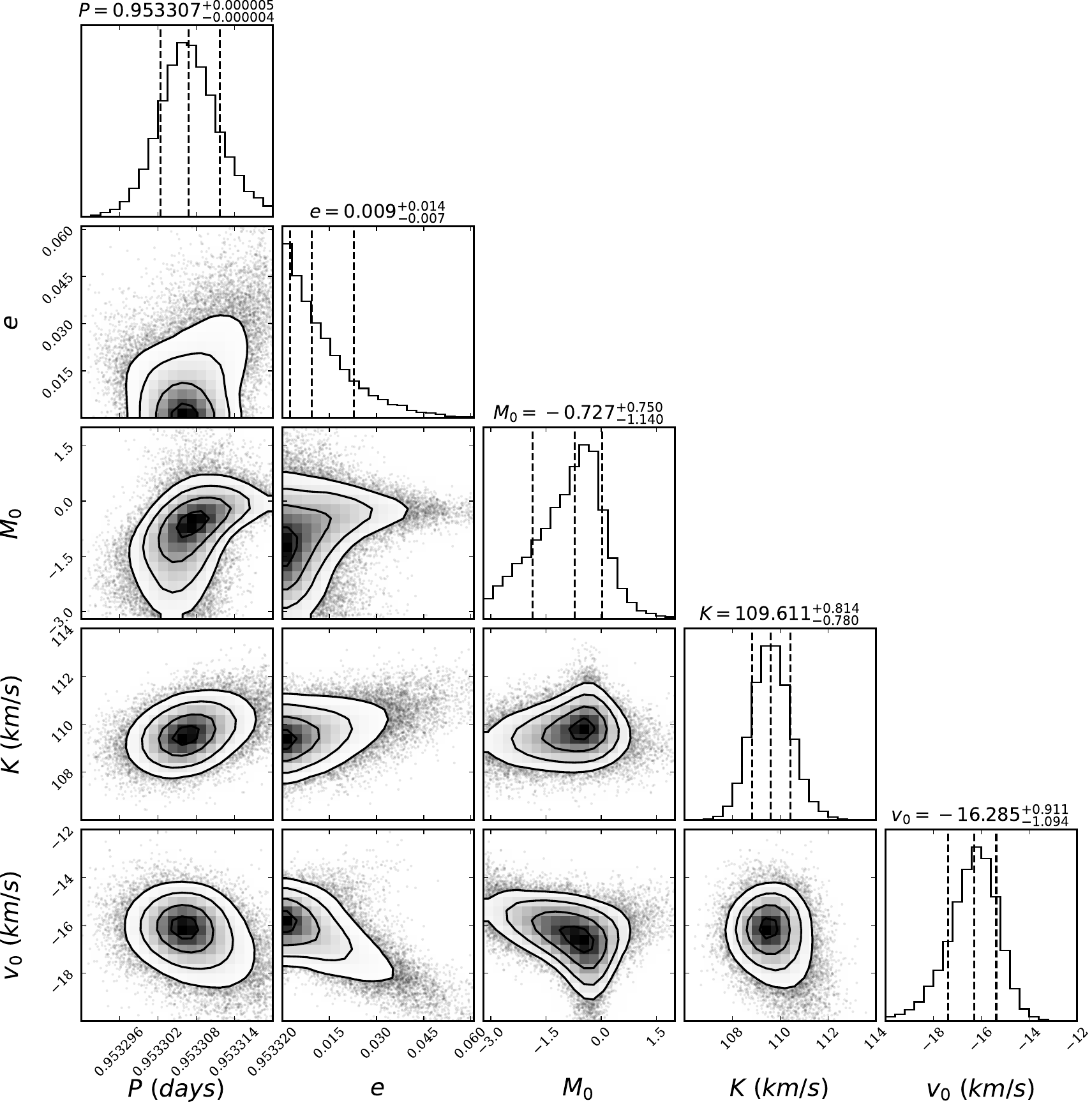}\\
    \includegraphics[width=\columnwidth]{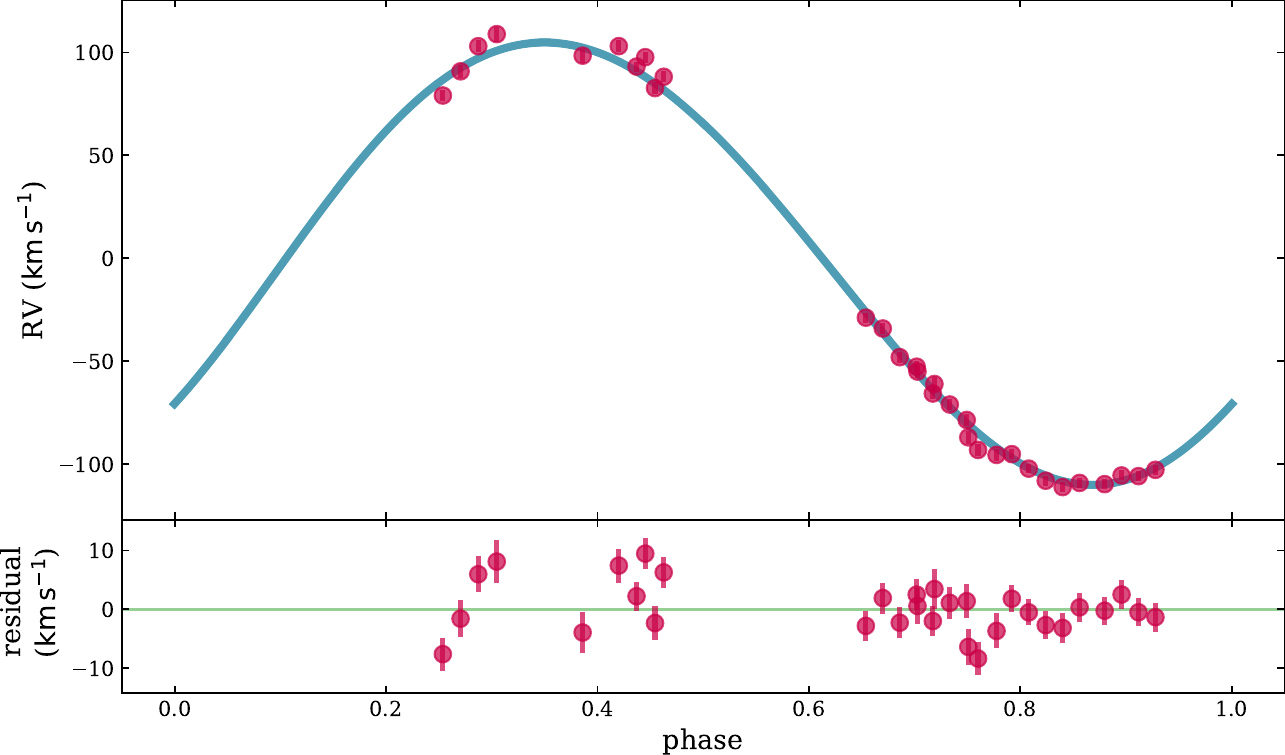}
    
    \caption{Top panel: The corner plot illustrates the distribution of orbital parameters for J2308, derived using the Joker. The parameters are labeled as the orbital period (P, in days),  the argument of pericenter ($\omega$, in radians), the mean anomaly at the reference time ($M_0$ in radians), the radial velocity semi-amplitude of the star ($K$, in $\rm km~s^{-1}$), and the center of mass velocity ($v_{0}$, in $\rm km~s^{-1}$). Bottom panel: Phase-folded radial velocities data (red dots) and the best-fit radial velocities curves (blue line) for J2308.}
    \label{figure1}
\end{figure}

The results of our fitting and the period-folded results are presented in  Figure~\ref{figure1}. 
We obtained a period $P=0.953307_{-0.000004}^{+0.000005}~\rm days$, the eccentricity $e=0.009_{-0.007}^{+0.014}$, an orbital semi-amplitude  
$K=109.611_{-0.780}^{+0.814}~\rm km~s^{-1}$, and calculated the mass function based on the fitting results for the radial velocity curve as $f(M_{2})=0.129 M_\odot$. 
Our result is consistent with that obtained in the previous study \cite{liu2024sample}, which reported a period of 0.95 days and a mass function of 0.128 $M_\odot$ for this system. We show the folded radial velocity curve based on the fitted parameters in the bottom panel of Figure~\ref{figure1}.

\subsection{ Estimation of Stellar Parameters}

We collected broadband spectral energy distribution (SED) fitting with the Python package astroARIADNE\footnote{\url{https://github.com/jvines/astroARIADNE}} to constrain the stellar parameters of the sources. The astroARIADNE is designed for automatically fitting broadband photometry to various stellar atmosphere models using Nested Sampling algorithms\citep{Vines2022}. 

\begin{figure}

    \includegraphics[width=\columnwidth]{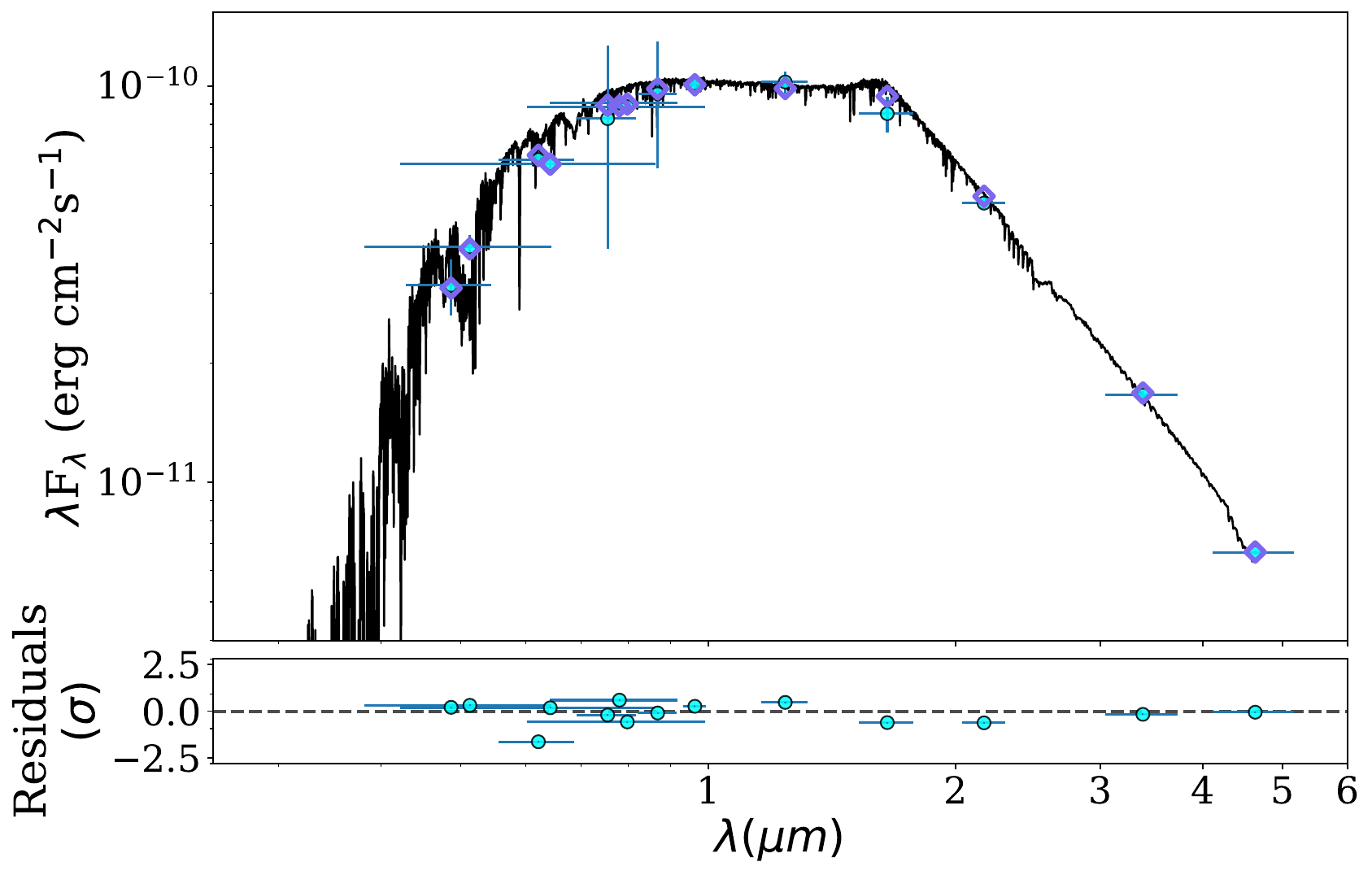}
    \caption{SED fitting of J2308. 
    The photometric data used for fitting (violet circles) are from Pan-STARR, Gaia, TESS, 2MASS and WISE. The black line represents the best-fit model.}
    \label{sed}
\end{figure}

We collect multiband photometric data including the Gaia \citep{brown2018gaia}, the Pan-STARRS \citep{chambers2016pan}, the Transiting Exoplanet Survey Satellite (TESS) \citep{ricker2015transiting}, the two micron all sky survey (2MASS) \citep{skrutskie2006two}, and the Wide-field Infrared Survey Explorer (WISE) \citep{wright2010wide}  to fit the SED. Additionally, we used stellar evolution models to estimate the mass of this source through the Python package isochrones\citep{morton2015isochrones}. By incorporating the best-fit SED parameters and photometric data, we calculated the interpolated mass of the source using the MESA Isochrones and Stellar Tracks (MIST) \citep{dotter2016mesa}.  
The Gaia DR3 parallax \citep{vallenari2023gaia}, as well as the extinction $A_{V}=0.22$, is also used. 
We plotted the SED fitting results in Figure~\ref{sed}, and the parameters derived from the fitting for this source are listed in Table~\ref{tsed}.

\begin{figure}

    \includegraphics[width=\columnwidth]{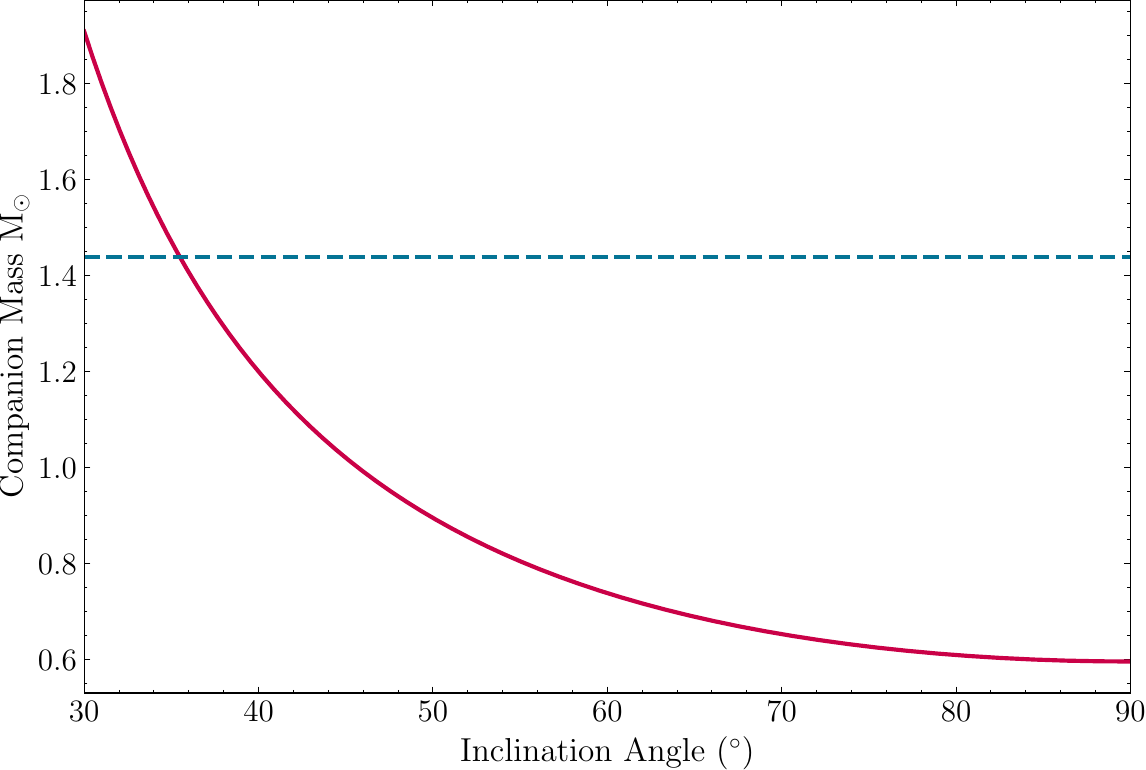}\\
    \includegraphics[width=\columnwidth]{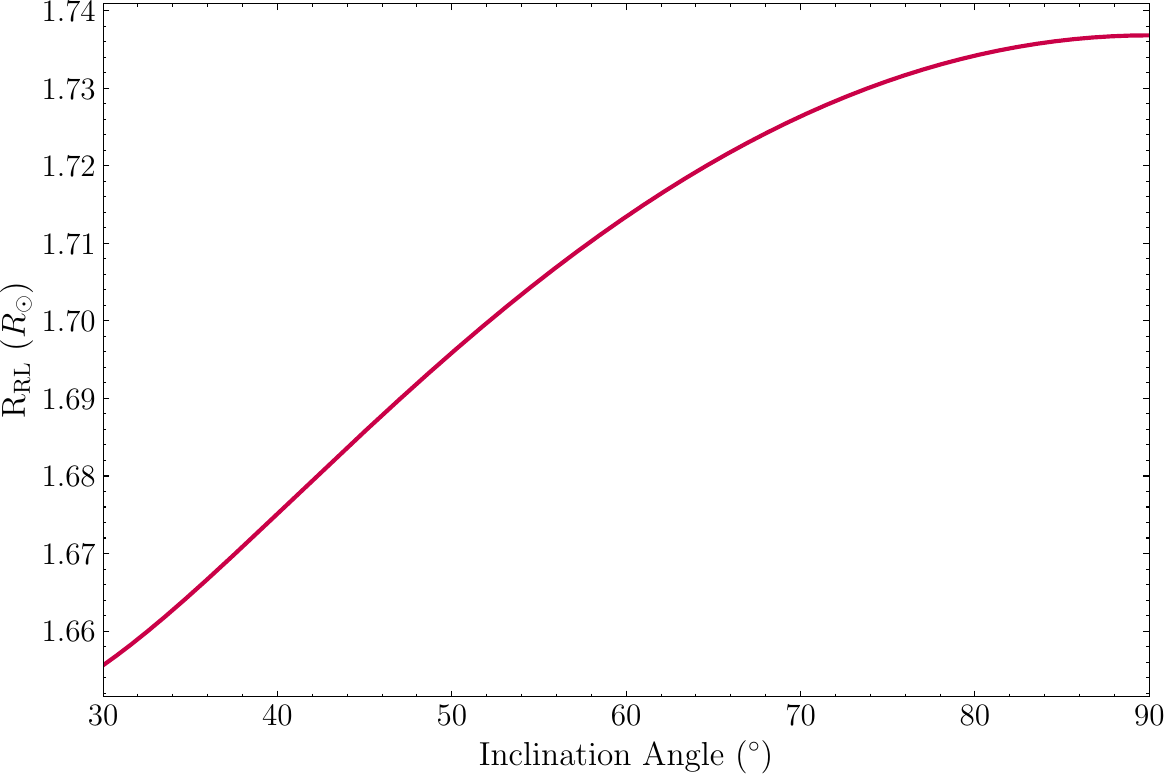}

    \caption{Top panel: the relationship between the mass of the invisible star and the system's inclination. The blue dashed line indicates the Chandrasekhar limit. Bottom panel: the relationship between the Roche lobe radius and the system's inclination.  }
    \label{mass}
\end{figure}

We adopt the isochrone mass as the visible star's mass to constrain the invisible star's mass. We use the mass function equation to estimate the mass of the invisible star at different inclination angles, and the results are presented in Figure~\ref{mass} top panel. 
This source is consistent with a white dwarf across a broad range of inclinations. The mass of the invisible star exceeds the Chandrasekhar limit only when the inclination is below $\sim 35$ degrees. 
In the most extreme case, assuming the orbital inclination of the binary system is uniformly distributed in $\cos~i$, the probability that its mass is less than the Chandrasekhar limit should be about 0.82.

\begin{table}
%\bc
\centering
%\begin{minipage}[]{100mm}
\caption{ Results of the stellar parameters of the visible star from SED fitting. A$_V$ is fixed for Gaia. }
  \renewcommand{\arraystretch}{0.85}
 \setlength{\tabcolsep}{4.5pt}
\begin{tabular}{cccc}
 \hline\noalign{\smallskip}

Parameter & Median & Upper &Lower \\

 \hline\noalign{\smallskip}

T$_\mathrm{eff}$ (K) & 4162.90 & 23.69 & 23.69  \\
log(g) & 4.54 & 0.07 & 0.08  \\
$\rm [Fe/H]$ (dex) &  0.16 & 0.05 &  0.04 \\
Distance (pc) & 154.10 & 0.49 & 0.44 \\
Radius (R$_\odot$) & 0.67 & 0.01 & 0.01  \\
A$_V$ & 0.22 & - & - \\
Isochrone Mass (M$_\odot$) & 0.68 & 0.03 & 0.02  \\

 \noalign{\smallskip}\hline
\label{tsed}
\end{tabular}

  \end{table}

Using the parameters mentioned above, we can calculate a simple solution for the Roche lobe radius. 
For simplicity, we assume an inclination angle of 90 degrees, which gives a Roche lobe radius upper limit of about $1.736~R_{\odot}$, according to the standard Roche lobe approximation \citep{eggleton1983approximations}:
\begin{equation}
    R_{RL} = \frac{0.49 q^{2/3}a}{0.6 q^{2/3} + \ln\left(1 + q^{1/3}\right) },
\end{equation}
where $a=\left(1 +q\right) P \left(1 - e^2\right)^{1/2} K/2\pi \sin i$. We also present the relationship between the calculated Roche lobe radius $ R_{RL} $ and the inclination angle $i$ in the bottom panel of Figure~\ref{mass}. From the figure, it is evident that the Roche lobe radius consistently exceeds 1.5 $R_{\odot}$, the radius of the visible star, approximately $R_{\odot}$  as obtained from the SED fitting, is significantly smaller than the Roche lobe radius. The visible star has not filled its Roche lobe, indicating that there is no ongoing mass transfer.
  
\subsection{ Light Curves and Spectra}
\begin{figure*}

    \includegraphics[width=0.9\textwidth]{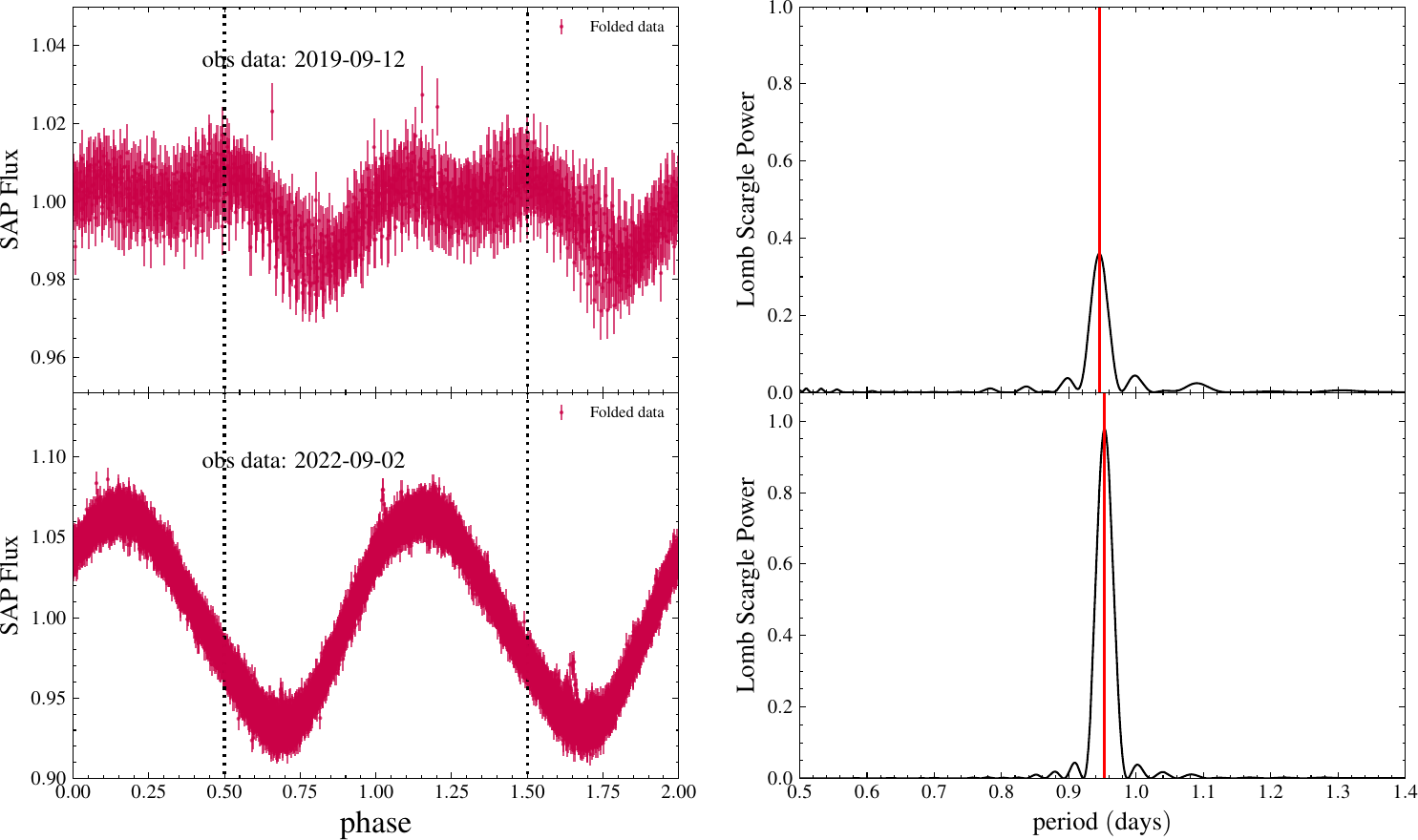}

    \caption{Left panel: the phase-folded TESS light curve. The top panel shows the 2019 observation data, while the bottom panel presents the 2022 observation data. In both panels, "obs data" refers to the start time of the observations.
    Right panel: The Lomb–Scargle periodograms corresponding to the observed data on the left panel. The red vertical line represents the period with the maximum power.}
    \label{tess}
\end{figure*}

TESS observed this source in both 2019 and 2022. Using the Lomb–Scargle algorithm to perform period searches on the obtained observational data, we found that the periods of the light curves were 0.9453 days and 0.9530 days, respectively. The period obtained from the light curve is very close to the period derived from the radial velocity. We plotted the phase-folded light curve in Figure ~\ref{tess}. 

\begin{figure*}

    \includegraphics[width=0.9\textwidth]{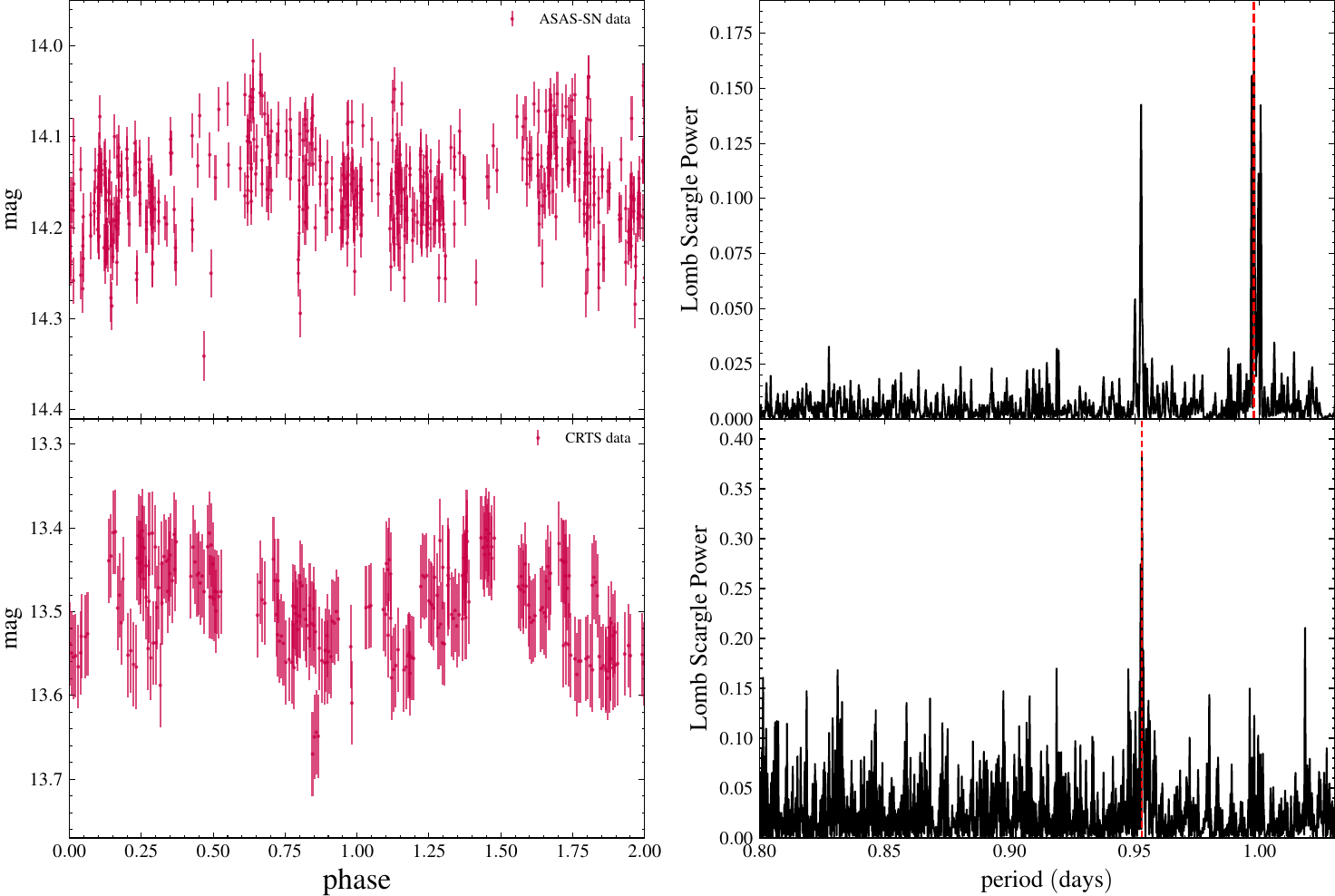}
    \caption{Left panel: top panel is the phase-folded light curve using ASAS-SN data. Bottom panel is the phase-folded light curve using CRTS data. Right panel: The Lomb–Scargle periodograms corresponding to the observed data on the left panel. The red vertical line represents the period with the maximum power. }
    \label{asas}
\end{figure*}

Additionally, we searched for this source in the All-Sky Automated Survey for Supernovae (ASAS-SN) database \cite{shappee2014man}, where it is listed as ASASSN-V J230854.10+355132.0 and classified as a Rotational Variable. The observation period was from July 6, 2018, to December 2, 2019, with a period of P = 0.9982674 days. We also conducted a periodicity search using data from the Catalina Real-Time Transient Survey (CRTS) \citep{drake2009first}, with the observation period ranging from June 16, 2005, to October 24, 2013, and the periodicity search resulted in a period of 0.9529 days.  We have plotted the folded results in Figure~\ref{asas}.
%In addition, we collected other photometric data, including the Zwicky %Transient Facility (ZTF; \citealt{bellm2019zwicky}), the All-Sky Automated %Survey for Supernovae (ASAS-SN; \cite{shappee2014man}), and the Wide-field %Infrared Survey Explorer(WISE; \citealt{wright2010wide}), to conduct %periodic searches on their light curves. 

\begin{figure}

    \includegraphics[width=\columnwidth]{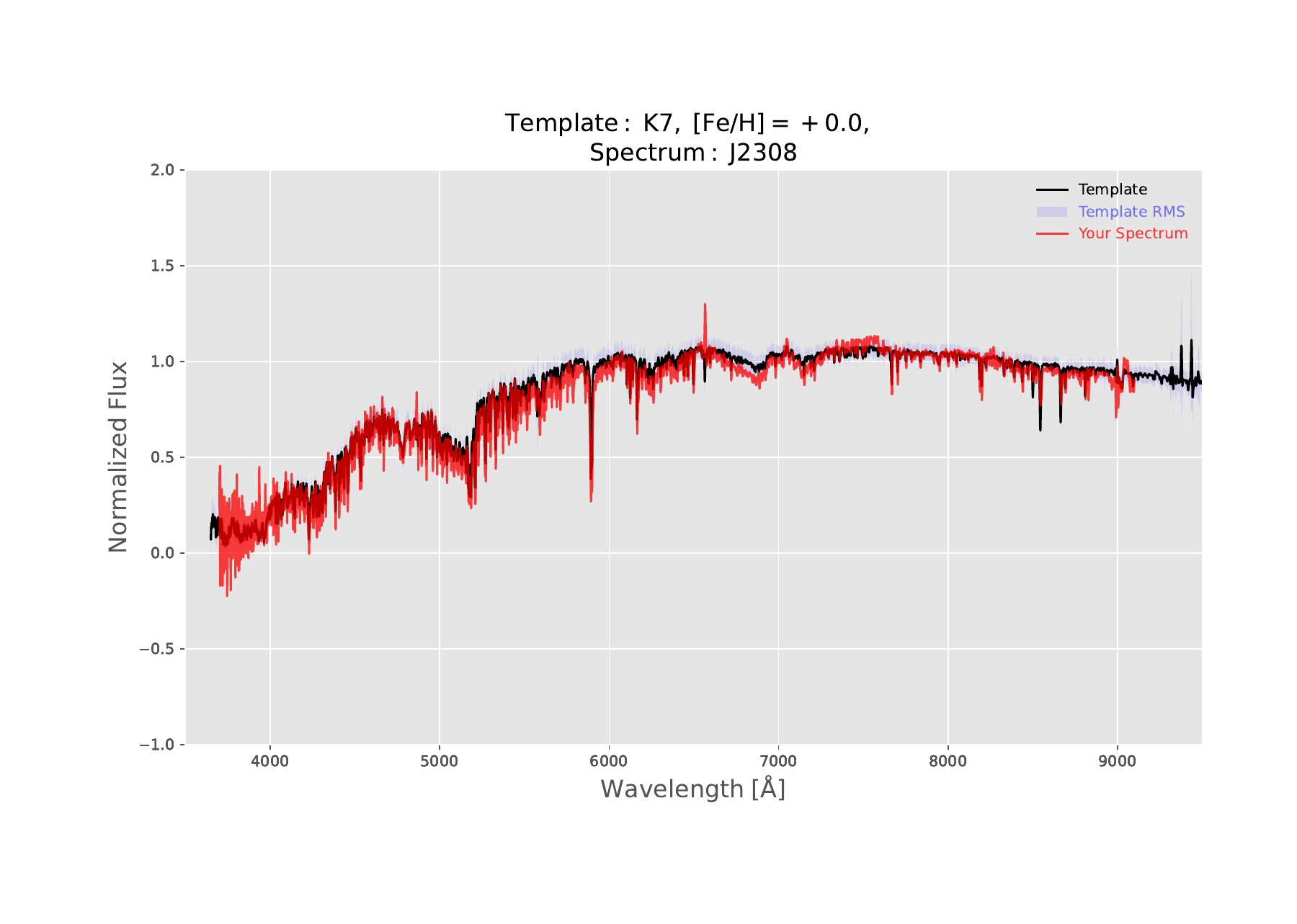}
    \caption{The fitting results using PyHammer. The black curve represents the observed LAMOST LRS spectrum, while the red curve shows the generated template spectrum. }
    \label{pyham}
\end{figure}

We searched the LAMOST DR11 LRS database for spectra of this source and found one spectrum from the observation on November 19, 2014. We fit the spectra of this source using a mixture of stellar models provided by the PyHammer package\footnote{\url{https://github.com/BU-hammerTeam/PyHammer}} \citep{kesseli2017empirical,roulston2020classifying}.  The spectral type derived from the fitting is K7 as shown in Figure~\ref{pyham}. This result indicates that the system can be well-fitted with a single star template, and it is not a spectroscopic binary with double-lined spectra. 

\begin{figure}

    \includegraphics[width=\columnwidth]{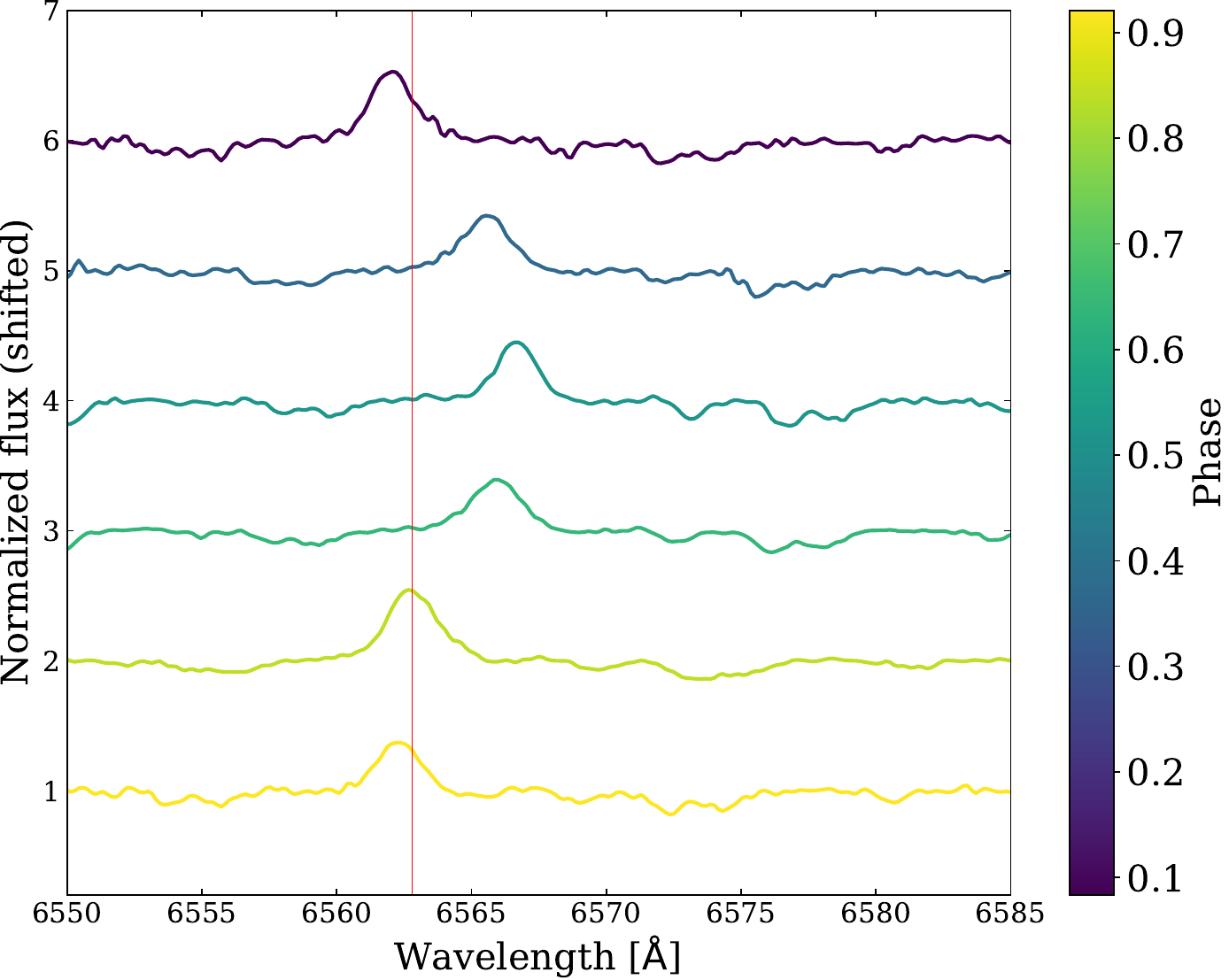}   
    \caption{the phase-dependent $\rm H_{\alpha}$ emission from J2308 in the LAMOST MRS spectra. The position marked by the red line corresponds to a wavelength of 6562.8 \AA. 
 }
    \label{halpha}
\end{figure}

In the LAMOST MRS, this source exhibits a prominent H$\alpha$ emission line. We selected six observations to display in Figure~\ref{halpha}. From the figure, it is evident that during one orbital period, the H$\alpha$ emission line shows periodic redshifts and blueshifts. Moreover, the emission line consistently exhibits a stable single-peaked structure throughout.
\section{Discussion} \label{sec4}
We identified a white dwarf binary candidate, J2308, using radial velocity search methods from the LAMOST MRS stellar catalog. This source was also identified as a single-lined spectroscopic binary in the previous work \citep{liu2024sample} during their study of LAMOST DR10, but it was only preliminarily confirmed as a binary star system. In our work, we have conducted a detailed analysis of this source, and identified the nature as a white dwarf binary.

\subsection{Binary Properties}

Using the Joker algorithm, we derived the orbital solution for J2308, obtaining an orbital period, mass function, and other parameters that are consistent with those previously reported. We performed SED fitting using photometric data across multiple bands and obtained stellar parameters indicating that the visible star is a K-type star with a mass of $0.68_{-0.02}^{+0.03} M_{\odot}$, a radius of $0.67 \pm 0.01 R_{\odot}$, an effective temperature of $4162.90\pm 23.69$ K, and a surface gravity of $4.54_{-0.07}^{+0.08}$. Furthermore, we used these parameters to constrain the mass of the invisible star and found that, for orbital inclinations greater than $\sim 35 ^\circ$, its mass is less than the Chandrasekhar limit. Even in extreme cases, the probability that the mass is below the Chandrasekhar limit is 0.81. If the compact object in this system is a neutron star, calculating with the current minimum mass of a neutron star with about 1.17 $M_{\odot}$ \citep{suwa2018minimum}, the mass function indicates an inclination angle of $\sim 40^\circ$.

Considering a more general situation, the systems for which we can observe larger changes in radial velocity should correspond to binary systems with high inclination angles. For example, in the LAMOST J112306.9+400736, the inclination angle obtained is $73^{+1.8}_{-1.5} \, ^\circ$ \citep{yi2022dynamically}. The inclination angles for the other three systems based on LAMOST data are also large\citep{qi2023searching}: for J034813, the inclination is approximately $55.36^\circ$; in the case of J063350, the inclination is nearly edge-on, estimated at approximately $84.81^\circ$; for J064850, the inclination is approximately $50.82^\circ$. This indicates that the invisible star in this binary system is most likely a white dwarf based on the large variations in radial velocity. 

Generally, the inclination angle can be constrained by fitting the ellipsoidal variations observed in the high-cadence TESS data. The folded light curve in Figure~\ref{tess} obtained from only two TESS observation data clearly shows that this source exhibits significant variability. The possible origin of such complex features is attributed to strong stellar activity. This is supported by the LAMOST spectra, as discussed in the following subsection. This leads to increased model complexity during the fitting process, and we did not achieve a good fitting result, which necessitates further detailed research. Additionally, by calculating its Roche lobe radius, we found that the visible star has not filled its Roche lobe, indicating that there is no ongoing mass transfer. The two stars in the system can be regarded as evolving independently. This could also result in very weak ellipsoidal modulation for the system. 
=In addition, we used PyHammer to fit the LRS. This result shows that the system can be well fitted with a single star template, suggesting that it is not a spectroscopic binary with double-lined spectra. 

%Previous studies using LAMOST LRS have similarly confirmed numerous samples exhibiting spectral characteristics of K/M-dwarf + white dwarf systems \citep{mu2022compact}. 

\subsection{Light Curve}

We used long-term observations from TESS in 2019 and 2022 to fold the light curves, and we found that the period matches the dynamic period. 
In the TESS 2019 observations, the periodic results showed that the system exhibits the characteristic features of an ellipsoidal modulation within the period. 
This modulation feature is similar to those observed in several other sources, such as the two white dwarf binaries Gaia DR3 4014708864481651840 and 5811237403155163520 \cite{rowan2024hidden}. The TESS light curves reveal ellipsoidal modulations caused by the tidal distortion of the K-dwarf star by a nearby stellar companion. 

However, in the 2022 observations, we did not detect such modulation. Instead, the light curves exhibited a very clear periodic variation see Figure~\ref{tess} bottom panel. We also used data from ASAS-SN and CRTS, which, when folded, revealed similar periodic features. Neither dataset showed ellipsoidal modulation, as illustrated in Figure~\ref{asas}. 
Additionally, this source has been classified by ASAS-SN as a rotational variable\citep{jayasinghe2019asas}. A similar situation occurred in the study of LAMOST J120802.64+311103.9, which was also classified as a variable star. However, the fitting results from PyHammer indicate that this system is a binary system containing a white dwarf \citep{mu2022compact,rowan2024hidden}.

In the LAMOST MRS spectra, J2308 exhibits variable H$\alpha$ emission. Typically, H$\alpha$ emission could originate from a combination of chromospheric activity and/or mass transfer.
In many previous studies, H$\alpha$ emission lines have also been discovered in such systems.
In the LAMOST J235456.73+335625.9 system, a strong H$\alpha$ emission line is also observed \citep{zheng2023nearest,tucker2024weighing}. In this system, the H$\alpha$ emission observed in the LAMOST spectra suggests that the UV excess is caused by chromospheric activity from the active K-dwarf. In 2MASS J15274848+3536572, the detailed analysis discovered that the LAMOST MRS spectrum reveals H$\alpha$ emission with a wider range and an additional peak, suggesting the presence of an accretion disk \cite{lin2023x}.  
The non-accreting neutron star–M dwarf binary candidate LAMOST J112306.9+400736 \cite{yi2022dynamically}, also exhibits H$\rm \alpha$ emission lines. They found that these emission lines are comoving with the M dwarf, suggesting that they originate from its hot chromosphere. Additionally,  double-peaked emission is commonly observed in accreting compact object binaries \cite{swihart2022new}. 

For J2308, our calculations show that its Roche lobe radius is significantly larger than its physical radius, indicating that there is no mass exchange in this system. The H$\alpha$ emission line does not exhibit broad features or additional components, suggesting that this emission likely originates from chromospheric activity. Analysis of the light curves in Figure~\ref{tess} bottom panel and ~\ref{asas} reveals that the surface of the K-type star displays strong hot spots. However, during specific observations, such as the TESS observation in 2019 (see Figure~\ref{tess}), the intensity of these hot spots diminishes, making the ellipsoidal modulation caused by binary motion more prominent in the light curve. At other times, the hot spots dominate the light curve, obscuring the modulation effects and leading to its previous classification as a rotational variable \cite{jayasinghe2019asas}. To summarize, in 2019, the star's activity may have been relatively subdued, allowing for the detection of more regular ellipsoidal modulation features. Subsequently, the star likely entered a phase of heightened activity. Additionally, due to the large Roche lobe radius, the ellipsoidal modulation itself is not prominent, which resulted in the light curve variations exhibiting characteristics typical of a rotational variable. 

\subsection{Binary System Constraints and Kinematic Analysis}
We previously presented the relationship between the invisible star and the orbital inclination. Here, we attempt to make further constraints using the TESS light curve from 2019, which exhibits ellipsoidal variations. We utilize the PHOEBE 2.4 \citep{conroy2020physics} software\footnote{\url{https://github.com/phoebe-project/phoebe2}} to fit the light curves. Specifically, we aim to model the light curves as spotted ellipsoidal variables (ELLs) using PHOEBE to derive the mass ratio and binary inclination. 

For our PHOEBE models, we treat the invisible star as a dark companion by fixing it to be small ($ R_2 = 3 \times 10^{-6} \, R_{\odot} $) and cold ($ T_{\text{eff, 2}} = 300 \, \text{K} $)\citep{jayasinghe2022giraffe,lin2023x}. We do not include the effects of irradiation or reflection and set distortion method=none to model the invisible star as an object without any contributions to flux\citep{qi2023searching,rowan2024hidden}. We also fix the eccentricity ($e = 0$) based on the RV fit. The spot on a visible star has an independent latitude ($\theta_s$), longitude ($\varphi_s$), angular size ($R_s$) and temperature ($T_s$), parameterized as a relative temperature ($T_s / T_{\text{eff}}$). We run this model for 10 000 iterations with 16 walkers. 

\begin{figure}
    \includegraphics[width=\columnwidth]{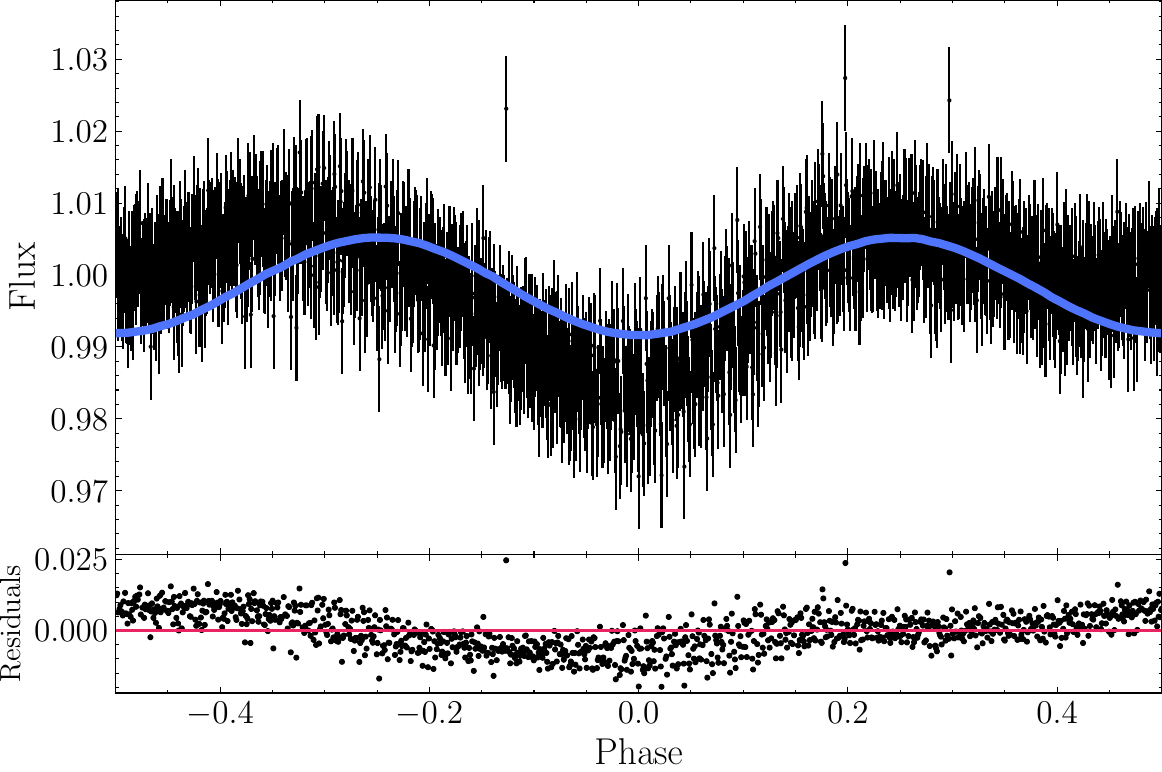}
    \includegraphics[width=\columnwidth]{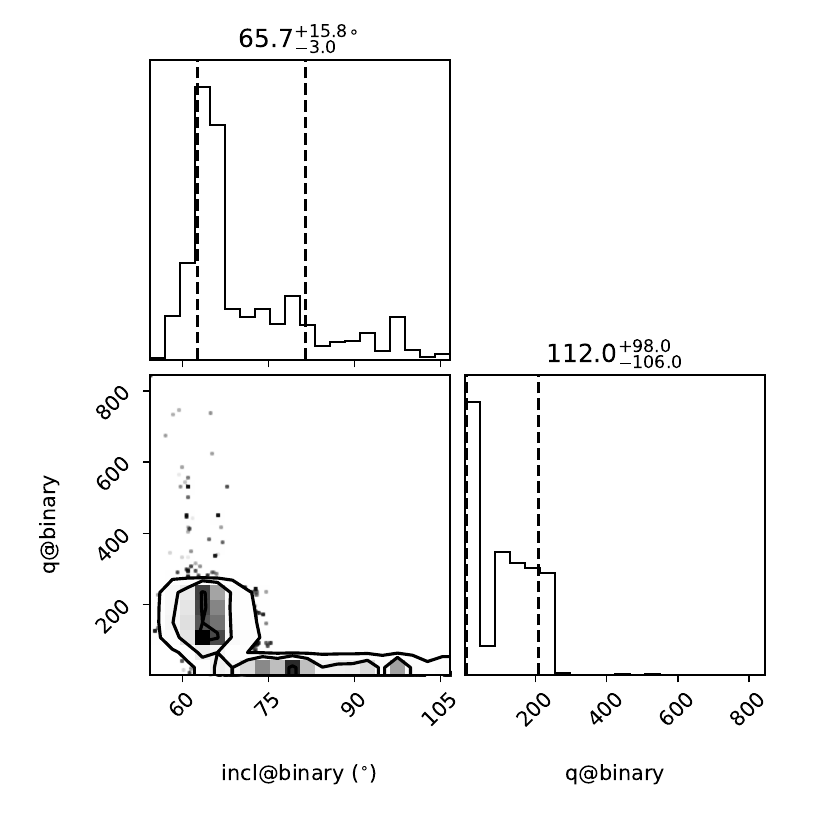}
    \caption{Top panel: Fitting results and residuals using the spot-free model. The black points represent the observational data after phase folding, with the phase 0 point set at the time of minimum flux. The blue line represents the fitted model, and the red line indicates the position where the residuals are zero.
Bottom panel: Parameter distributions of the light curve fitting.  }
    \label{nospot}
\end{figure}

\begin{figure}

    \includegraphics[width=\columnwidth]{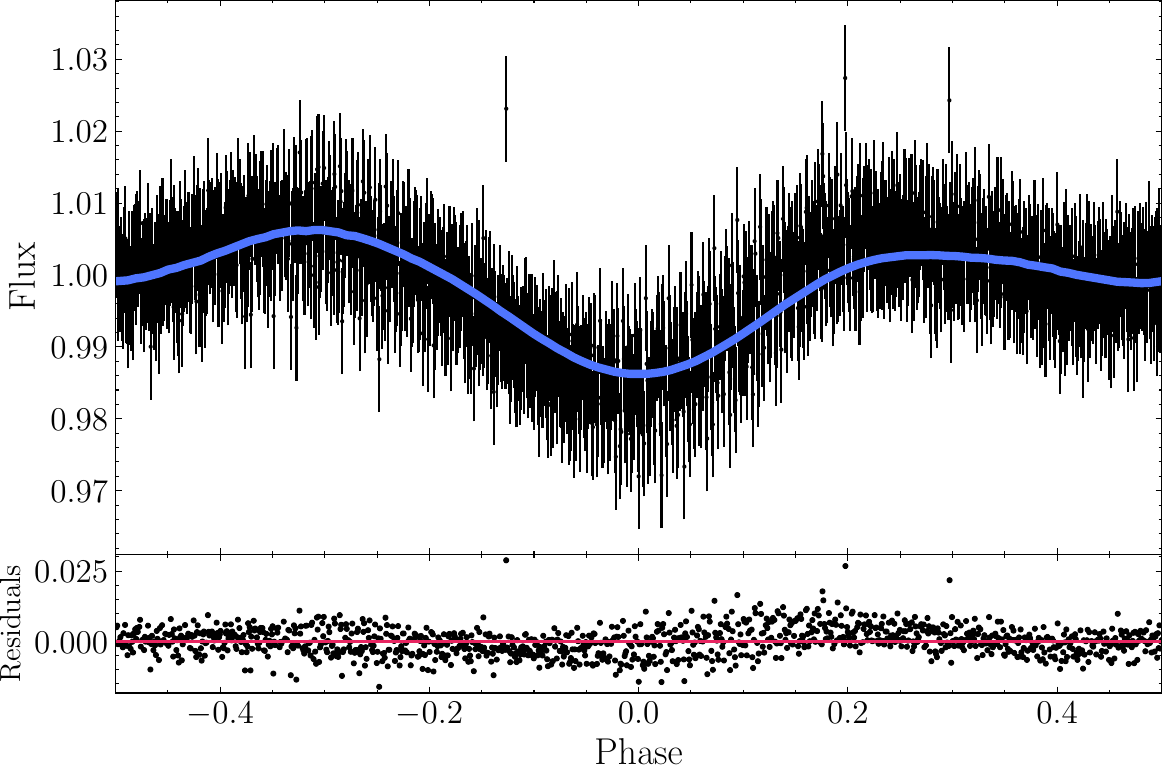}
    \includegraphics[width=\columnwidth]{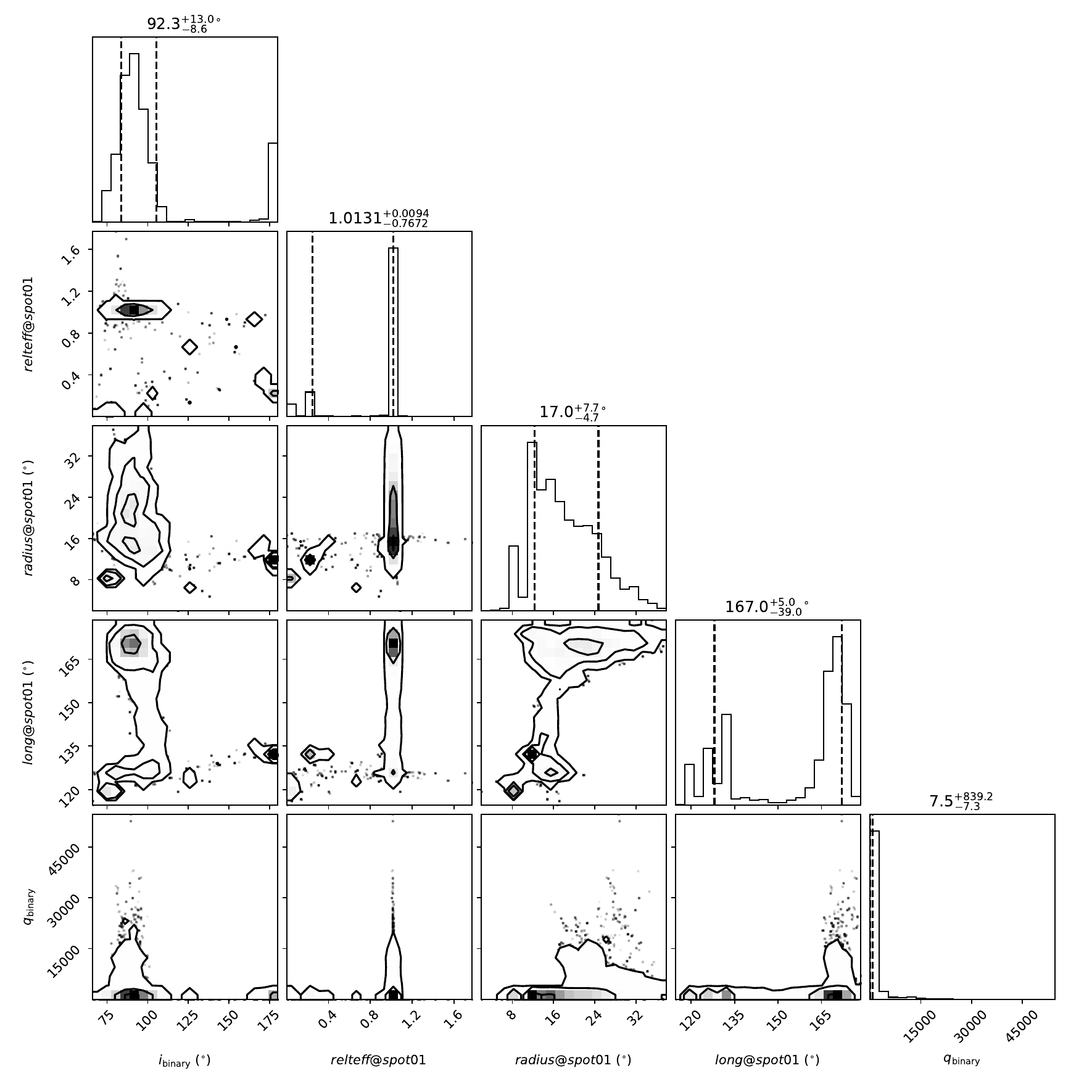}
    \caption{Top panel: Fitting results and residuals using the single spot model. The black points represent the observational data after phase folding, with the phase 0 point set at the time of minimum flux. The blue line represents the fitted model, and the red line indicates the position where the residuals are zero.
Bottom panel: Parameter distributions of the light curve fitting. }
    \label{spot}
\end{figure}

We presented the fitting results of the spotless model and single spot model in Figures~\ref{nospot} and ~\ref{spot}, respectively, where the black points represent the data points, and the blue line corresponds to the model. As shown in Figures~\ref{nospot}, there  is a noticeable discrepancy between the spotless model and the observations, suggesting that the model fails to accurately capture the observed light curve and does not offer strong constraints on the parameters. In Figures~\ref{spot}, despite the single spot model provides a better fit, accurately capturing the first peak, but it does not adequately describe the second peak of the observed data. Moreover, the parameter distributions from the fit suggest that while the model provides reasonable estimates for the spot's parameters, it does not place strong constraints on the physical parameters of the entire system.

Although the observed light curve could not be fully reproduced, we found that both models indicate a preference for a high inclination angle ($>$60 degrees). By relating this inclination to the mass-inclination relationship, we determined that within this range, the mass of the unseen companion is constrained to be less than $0.8 ~M_\odot$. This result provides rough constraints on the nature of the invisible star, suggesting that it could potentially be a white dwarf. 

We believe that the light curve model for this source requires considering additional factors or the inclusion of multiple spots in the fitting process. Furthermore, it is necessary to account for the effects of different spot distributions on the light curve to achieve a more accurate representation of the system.

\begin{figure}
    \includegraphics[width=\columnwidth]{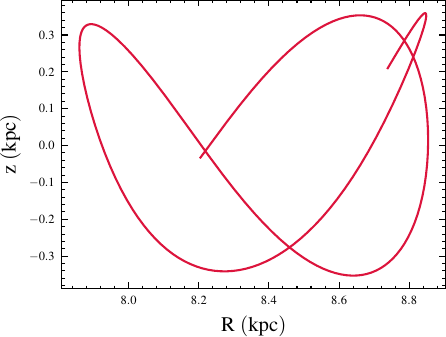}
    \includegraphics[width=\columnwidth]{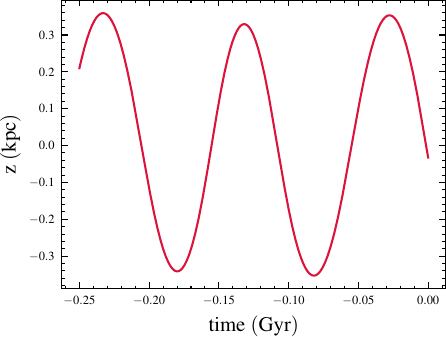}
    
    \caption{The top panel shows the variation of the source's position in Galactocentric cylindrical coordinates. The bottom panel illustrates the relationship between z and time during the integration period. }
    \label{or}
\end{figure}

The kinematics of white dwarf and neutron star binaries are expected to differ. Neutron stars often receive natal kicks during supernovae, affecting their motion in the Galaxy. This system is likely to exhibit atypical Galactic orbits if it contains a neutron star. Natal kicks are imparted to neutron stars during supernova explosions and are typically on the order of $\rm 100-500~ km~s^{-1}$, and they can reach up to $\rm \sim 1000~ km~s^{-1}$ in extreme cases\citep{hobbs2005statistical}. 
However, some studies suggest that the scalar speeds of older neutron stars provide little information about the kicks they received at birth \citep{disberg2024deceleration}.  We used the Gaia DR3 parallax, proper motion, and the center-of-mass velocity derived from the radial velocities to estimate the trajectory of this source within the Galaxy. we use galpy\citep{bovy2015galpy} to integrate the backward orbit over a total time of 250 Myr with a time step of 0.001 Myr in MWPotential2014 potential. We adopted the currently widely used parameter values for the local standard of rest  velocity, as well as the solar's position and motion parameters\citep{liu2024exploration}. We present the calculated orbital trajectory in the Figure~\ref{or}. The calculated orbits are consistent with the thin disc population, remaining within approximately 300 pc of the Galactic mid-plane\citep{yan2019chemical,zhu2021element}  It should be noted that our calculations only consider a simple Galactic gravitational potential model and do not account for the effects of dynamical friction. For a more detailed dynamical analysis, more complex potential models can be introduced\citep{liao2023hypervelocity}. 

\section{Conclusions}
\label{sec5}
 We presented a binary system containing a white dwarf candidate discovered using LAMOST MRS data based on dynamical method. This sample of binary systems with the compact objects is still relatively rare, and their some unique properties are also interesting for binary studies and evolution. The main results of the study can be summarized as follows: 

\begin{itemize}
\item By analyzing the Radial Velocity data from the LAMOST MRS, we determined the orbital periods, fitted the radial velocity curve, and computed the mass function. The derived orbital period is approximately 0.953 days, and the mass function is 0.129 $M_\odot$. 

\item We used SED fitting to estimate the stellar parameters, and the results indicate that the visible star has an effective temperature of approximately 4100 K, a mass of $0.68_{-0.02}^{+0.03} M_{\odot}$, and a radius of $0.67_{-0.01}^{+0.01 } R_{\odot}$. 

\item We obtained the relationship between the mass of the invisible object and the system's inclination angle, and we also calculated the relationship between the Roche lobe radius and the inclination angle. The results show that the mass of the invisible star is below the Chandrasekhar limit when the inclination angle exceeds $\sim 35$ degrees. The PHOEBE fitting results suggest the inclination angle of this system $>$60 degrees, thus the mass of the invisible star is estimated to be less than approximately $0.8~M_{\odot}$, then we can conclude that the invisible star in this system is a white dwarf. Additionally, the Roche lobe radius is larger than the physical radius of the visible star, indicating that no mass transfer occurs in this system. The larger Roche lobe radius also results in a very weak ellipsoidal modulation effect in this system.

\item We also obtained light curves from the TESS, ASAS-SN, and CRTS surveys for this target. After applying the Lomb–Scargle algorithm to phase-fold the light curves, we found periods around 0.95 days. However, only the observations from TESS in 2019 exhibited a noticeable ellipsoidal modulation. Combined with the prominent $\rm H_{\alpha}$ emission line observed in the LAMOST MRS spectrum we infer that the surface of the visible star exhibits strong thermal spots. Coupled with the system's weak ellipsoidal modulation effect, these factors lead to the system exhibiting characteristics of rotational variability. 

\end{itemize}

\section*{Acknowledgements}
We are grateful to the referee for the useful comments to improve the manuscript. This work is supported by the National Key Research and Development Program of China (Grants No. 2023YFA1607901 and 2021YFA0718503), the NSFC (12133007), the Youth Program of Natural Science Foundation of Hubei Province (2024AFB386) and the Postdoctoral Fellowship Program (Grade C) of China Postdoctoral Science Foundation (Grant No. GZC20241282). This paper uses the data from the LAMOST survey. Guoshoujing Telescope (the Large Sky Area Multi-Object Fiber Spectroscopic Telescope LAMOST) is a National Major Scientific Project built by the Chinese Academy of Sciences. Funding for the project has been provided by the National Development and Reform Commission.
%% The Appendices part is started with the command \appendix;
%% appendix sections are then done as normal sections
%\appendix
%\section{Example Appendix Section}
%\label{app1}

%Appendix text.

%% For citations use: 
%%       \cite{<label>} ==> [1]

%%
%Example citation, See \cite{lamport94}.

%% If you have bib database file and want bibtex to generate the
%% bibitems, please use
%%
%%  \bibliographystyle{elsarticle-num} 
%%  \bibliography{<your bibdatabase>}

%% else use the following coding to input the bibitems directly in the
%% TeX file.

%% Refer following link for more details about bibliography and citations.
%% https://en.wikibooks.org/wiki/LaTeX/Bibliography_Management
 \bibliographystyle{elsarticle-num} 
 \bibliography{elsarticle-template-num}

\begin{thebibliography}{10}
\expandafter\ifx\csname url\endcsname\relax
  \def\url#1{\texttt{#1}}\fi
\expandafter\ifx\csname urlprefix\endcsname\relax\def\urlprefix{URL }\fi
\expandafter\ifx\csname href\endcsname\relax
  \def\href#1#2{#2} \def\path#1{#1}\fi

\bibitem{mullally2009twins}
F.~Mullally, C.~Badenes, S.~E. Thompson, R.~Lupton, Twins: The two shortest period non-interacting double degenerate white dwarf stars, The Astrophysical Journal 707~(1) (2009) L51.

\bibitem{casewell2018first}
S.~Casewell, I.~Braker, S.~Parsons, J.~Hermes, M.~Burleigh, C.~Belardi, A.~Chaushev, N.~Finch, M.~Roy, S.~Littlefair, et~al., The first sub-70 min non-interacting wd--bd system: Epic212235321, Monthly Notices of the Royal Astronomical Society 476~(1) (2018) 1405--1411.

\bibitem{jayasinghe2023search}
T.~Jayasinghe, D.~Rowan, T.~A. Thompson, C.~Kochanek, K.~Stanek, A search for compact object companions to high mass function single-lined spectroscopic binaries in gaia dr3, Monthly Notices of the Royal Astronomical Society 521~(4) (2023) 5927--5939.

\bibitem{duchene2013stellar}
G.~Duch{\^e}ne, A.~Kraus, Stellar multiplicity, Annual Review of Astronomy and Astrophysics 51~(1) (2013) 269--310.

\bibitem{whitworth2015majority}
A.~P. Whitworth, O.~Lomax, Are the majority of sun-like stars single?, Monthly Notices of the Royal Astronomical Society 448~(2) (2015) 1761--1766.

\bibitem{remillard2006}
R.~A. Remillard, J.~E. McClintock, X-{{Ray Properties}} of {{Black-Hole Binaries}}, Annual Review of Astronomy and Astrophysics 44~(1) (2006) 49--92.
\newblock \href {https://doi.org/10.1146/annurev.astro.44.051905.092532} {\path{doi:10.1146/annurev.astro.44.051905.092532}}.

\bibitem{rebassa2012post}
A.~Rebassa-Mansergas, A.~Nebot G{\'o}mez-Mor{\'a}n, M.~R. Schreiber, B.~G{\"a}nsicke, A.~Schwope, J.~Gallardo, D.~Koester, Post-common envelope binaries from sdss--xiv. the dr7 white dwarf--main-sequence binary catalogue, Monthly Notices of the Royal Astronomical Society 419~(1) (2012) 806--816.

\bibitem{chen2023binary}
X.~Chen, Z.~Liu, Z.~Han, Binary stars in the new millennium, Progress in Particle and Nuclear Physics (2023) 104083.

\bibitem{cui2012large}
X.-Q. Cui, Y.-H. Zhao, Y.-Q. Chu, G.-P. Li, Q.~Li, L.-P. Zhang, H.-J. Su, Z.-Q. Yao, Y.-N. Wang, X.-Z. Xing, et~al., The large sky area multi-object fiber spectroscopic telescope (lamost), Research in Astronomy and Astrophysics 12~(9) (2012) 1197.

\bibitem{zhao2012lamost}
G.~Zhao, Y.-H. Zhao, Y.-Q. Chu, Y.-P. Jing, L.-C. Deng, Lamost spectral survey—an overview, Research in Astronomy and Astrophysics 12~(7) (2012) 723.

\bibitem{gu2019method}
W.-M. Gu, H.-J. Mu, J.-B. Fu, L.-L. Zheng, T.~Yi, Z.-R. Bai, S.~Wang, H.-T. Zhang, Y.-J. Lei, Y.~Bai, et~al., A method to search for black hole candidates with giant companions by lamost, The Astrophysical Journal Letters 872~(2) (2019) L20.

\bibitem{liu2019wide}
J.~Liu, H.~Zhang, A.~W. Howard, Z.~Bai, Y.~Lu, R.~Soria, S.~Justham, X.~Li, Z.~Zheng, T.~Wang, et~al., A wide star--black-hole binary system from radial-velocity measurements, Nature 575~(7784) (2019) 618--621.

\bibitem{wang2024potential}
S.~Wang, X.~Zhao, F.~Feng, H.~Ge, Y.~Shao, Y.~Cui, S.~Gao, L.~Zhang, P.~Wang, X.~Li, et~al., A potential mass-gap black hole in a wide binary with a circular orbit, Nature Astronomy (2024) 1--9.

\bibitem{swihart2021discovery}
S.~J. Swihart, J.~Strader, E.~Aydi, L.~Chomiuk, K.~C. Dage, L.~Shishkovsky, Discovery of a new redback millisecond pulsar candidate: 4fgl j0940. 3--7610, The Astrophysical Journal 909~(2) (2021) 185.

\bibitem{yi2022dynamically}
T.~Yi, W.-M. Gu, Z.-X. Zhang, L.-L. Zheng, M.~Sun, J.~Wang, Z.~Bai, P.~Wang, J.~Wu, Y.~Bai, et~al., A dynamically discovered and characterized non-accreting neutron star--m dwarf binary candidate, Nature Astronomy 6~(10) (2022) 1203--1212.

\bibitem{el2021lamost}
K.~El-Badry, E.~Quataert, H.-W. Rix, D.~R. Weisz, T.~Kupfer, K.~J. Shen, M.~Xiang, Y.~Yang, X.~Liu, Lamost j0140355+ 392651: an evolved cataclysmic variable donor transitioning to become an extremely low-mass white dwarf, Monthly Notices of the Royal Astronomical Society 505~(2) (2021) 2051--2073.

\bibitem{li2022}
G.~Li, S.~Deheuvels, J.~Ballot, F.~Ligni{\`e}res, Magnetic fields of 30 to 100 {{kG}} in the cores of red giant stars, Nature 610~(7930) (2022) 43--46.
\newblock \href {https://doi.org/10.1038/s41586-022-05176-0} {\path{doi:10.1038/s41586-022-05176-0}}.

\bibitem{qi2023searching}
S.~Qi, W.-M. Gu, T.~Yi, Z.-X. Zhang, S.~Wang, J.~Liu, Searching for compact object candidates from lamost time-domain survey of four k2 plates, The Astronomical Journal 165~(5) (2023) 187.

\bibitem{zheng2023nearest}
L.-L. Zheng, M.~Sun, W.-M. Gu, T.~Yi, Z.-X. Zhang, P.~Wang, J.~Wang, J.~Wu, S.-S. Weng, S.~Wang, et~al., The nearest neutron star candidate in a binary revealed by optical time-domain surveys, Science China Physics, Mechanics \& Astronomy 66~(12) (2023) 129512.

\bibitem{rowan2024hidden}
D.~M. Rowan, T.~Jayasinghe, M.~A. Tucker, C.~Y. Lam, T.~A. Thompson, C.~S. Kochanek, N.~S. Abrams, B.~J. Fulton, I.~Ilyin, H.~Isaacson, et~al., A hidden population of massive white dwarfs: two spotted k+ wd binaries, Monthly Notices of the Royal Astronomical Society 529~(1) (2024) 587--603.

\bibitem{jayasinghe2021unicorn}
T.~Jayasinghe, K.~Stanek, T.~A. Thompson, C.~Kochanek, D.~Rowan, P.~Vallely, K.~Strassmeier, M.~Weber, J.~Hinkle, F.~Hambsch, et~al., A unicorn in monoceros: the 3 $m_{\odot}$ dark companion to the bright, nearby red giant v723 mon is a non-interacting, mass-gap black hole candidate, Monthly Notices of the Royal Astronomical Society 504~(2) (2021) 2577--2602.

\bibitem{jayasinghe2022giraffe}
T.~Jayasinghe, T.~A. Thompson, C.~Kochanek, K.~Stanek, D.~Rowan, D.~Martin, K.~El-Badry, P.~Vallely, J.~Hinkle, D.~Huber, et~al., The ‘giraffe’: discovery of a stripped red giant in an interacting binary with an~ 2 $m_{\odot}$ lower giant, Monthly Notices of the Royal Astronomical Society 516~(4) (2022) 5945--5963.

\bibitem{el2022unicorns}
K.~El-Badry, R.~Seeburger, T.~Jayasinghe, H.-W. Rix, S.~Almada, C.~Conroy, A.~M. Price-Whelan, K.~Burdge, Unicorns and giraffes in the binary zoo: stripped giants with subgiant companions, Monthly Notices of the Royal Astronomical Society 512~(4) (2022) 5620--5641.

\bibitem{camisassa2019evolution}
M.~E. Camisassa, L.~G. Althaus, A.~H. C{\'o}rsico, F.~C. De~Ger{\'o}nimo, M.~M.~M. Bertolami, M.~L. Novarino, R.~D. Rohrmann, F.~C. Wachlin, E.~Garc{\'\i}a-Berro, The evolution of ultra-massive white dwarfs, Astronomy \& Astrophysics 625 (2019) A87.

\bibitem{kleinman2012sdss}
S.~J. Kleinman, S.~O. Kepler, D.~Koester, I.~Pelisoli, V.~Pecanha, A.~Nitta, J.~E. d.~S. Costa, J.~Krzesinski, P.~Dufour, F.-R. Lachapelle, et~al., Sdss dr7 white dwarf catalog, The Astrophysical Journal Supplement Series 204~(1) (2012) 5.

\bibitem{tremblay2016field}
P.-E. Tremblay, J.~Cummings, J.~Kalirai, B.~G{\"a}nsicke, N.~Gentile-Fusillo, R.~Raddi, The field white dwarf mass distribution, Monthly Notices of the Royal Astronomical Society 461~(2) (2016) 2100--2114.

\bibitem{lomb1976least}
N.~R. Lomb, Least-squares frequency analysis of unequally spaced data, Astrophysics and space science 39 (1976) 447--462.

\bibitem{scargle1981studies}
J.~D. {Scargle}, {Studies in astronomical time series analysis. I - Modeling random processes in the time domain}, The Astrophysical Journal Supplement 45 (1981) 1--71.

\bibitem{price2017joker}
A.~M. Price-Whelan, D.~W. Hogg, D.~Foreman-Mackey, H.-W. Rix, The joker: A custom monte carlo sampler for binary-star and exoplanet radial velocity data, The Astrophysical Journal 837~(1) (2017) 20.

\bibitem{li2022binaries}
X.~Li, S.~Wang, X.~Zhao, Z.~Bai, H.~Yuan, H.~Zhang, J.~Liu, Binaries with possible compact components discovered from the lamost time-domain survey of four k2 plates, The Astrophysical Journal 938~(1) (2022) 78.

\bibitem{zhang2022spectroscopic}
B.~Zhang, Y.-J. Jing, F.~Yang, J.-C. Wan, X.~Ji, J.-N. Fu, C.~Liu, X.-B. Zhang, F.~Luo, H.~Tian, et~al., The spectroscopic binaries from the lamost medium-resolution survey. i. searching for double-lined spectroscopic binaries with a convolutional neural network, The Astrophysical Journal Supplement Series 258~(2) (2022) 26.

\bibitem{liu2024sample}
H.-B. Liu, W.-M. Gu, Z.-X. Zhang, T.~Yi, J.-Z. Liu, M.~Sun, A sample of compact object candidates in single-lined spectroscopic binaries from lamost medium resolution survey, arXiv preprint arXiv:2405.09825 (2024).

\bibitem{Vines2022}
J.~I. {Vines}, J.~S. {Jenkins}, {ARIADNE: Measuring accurate and precise stellar parameters through SED fitting}, Monthly Notices of the Royal Astronomical Society (Apr. 2022).
\newblock \href {http://arxiv.org/abs/2204.03769} {\path{arXiv:2204.03769}}, \href {https://doi.org/10.1093/mnras/stac956} {\path{doi:10.1093/mnras/stac956}}.

\bibitem{brown2018gaia}
A.~Brown, A.~Vallenari, T.~Prusti, J.~De~Bruijne, C.~Babusiaux, C.~Bailer-Jones, M.~Biermann, D.~W. Evans, L.~Eyer, F.~Jansen, et~al., Gaia data release 2-summary of the contents and survey properties, Astronomy \& astrophysics 616 (2018) A1.

\bibitem{chambers2016pan}
K.~C. Chambers, E.~Magnier, N.~Metcalfe, H.~Flewelling, M.~Huber, C.~Waters, L.~Denneau, P.~Draper, D.~Farrow, D.~Finkbeiner, et~al., The pan-starrs1 surveys, arXiv preprint arXiv:1612.05560 (2016).

\bibitem{ricker2015transiting}
G.~R. Ricker, J.~N. Winn, R.~Vanderspek, D.~W. Latham, G.~{\'A}. Bakos, J.~L. Bean, Z.~K. Berta-Thompson, T.~M. Brown, L.~Buchhave, N.~R. Butler, et~al., Transiting exoplanet survey satellite, Journal of Astronomical Telescopes, Instruments, and Systems 1~(1) (2015) 014003--014003.

\bibitem{skrutskie2006two}
M.~Skrutskie, R.~Cutri, R.~Stiening, M.~Weinberg, S.~Schneider, J.~Carpenter, C.~Beichman, R.~Capps, T.~Chester, J.~Elias, et~al., The two micron all sky survey (2mass), The Astronomical Journal 131~(2) (2006) 1163.

\bibitem{wright2010wide}
E.~L. Wright, P.~R. Eisenhardt, A.~K. Mainzer, M.~E. Ressler, R.~M. Cutri, T.~Jarrett, J.~D. Kirkpatrick, D.~Padgett, R.~S. McMillan, M.~Skrutskie, et~al., The wide-field infrared survey explorer (wise): mission description and initial on-orbit performance, The Astronomical Journal 140~(6) (2010) 1868.

\bibitem{morton2015isochrones}
T.~D. Morton, isochrones: Stellar model grid package, Astrophysics Source Code Library (2015) ascl--1503.

\bibitem{dotter2016mesa}
A.~Dotter, Mesa isochrones and stellar tracks (mist) 0: methods for the construction of stellar isochrones, The Astrophysical Journal Supplement Series 222~(1) (2016) 8.

\bibitem{vallenari2023gaia}
A.~Vallenari, A.~G. Brown, T.~Prusti, J.~H. De~Bruijne, F.~Arenou, C.~Babusiaux, M.~Biermann, O.~L. Creevey, C.~Ducourant, D.~W. Evans, et~al., Gaia data release 3-summary of the content and survey properties, Astronomy \& Astrophysics 674 (2023) A1.

\bibitem{eggleton1983approximations}
P.~P. Eggleton, Approximations to the radii of roche lobes, Astrophysical Journal, Part 1 (ISSN 0004-637X), vol. 268, May 1, 1983, p. 368, 369. 268 (1983) 368.

\bibitem{shappee2014man}
B.~J. Shappee, J.~Prieto, D.~Grupe, C.~Kochanek, K.~Stanek, G.~De~Rosa, S.~Mathur, Y.~Zu, B.~Peterson, R.~Pogge, et~al., The man behind the curtain: X-rays drive the uv through nir variability in the 2013 active galactic nucleus outburst in ngc 2617, The Astrophysical Journal 788~(1) (2014) 48.

\bibitem{drake2009first}
A.~Drake, S.~Djorgovski, A.~Mahabal, E.~Beshore, S.~Larson, M.~Graham, R.~Williams, E.~Christensen, M.~Catelan, A.~Boattini, et~al., First results from the catalina real-time transient survey, The Astrophysical Journal 696~(1) (2009) 870.

\bibitem{kesseli2017empirical}
A.~Y. Kesseli, A.~A. West, M.~Veyette, B.~Harrison, D.~Feldman, J.~J. Bochanski, An empirical template library of stellar spectra for a wide range of spectral classes, luminosity classes, and metallicities using sdss boss spectra, The Astrophysical Journal Supplement Series 230~(2) (2017) 16.

\bibitem{roulston2020classifying}
B.~R. Roulston, P.~J. Green, A.~Y. Kesseli, Classifying single stars and spectroscopic binaries using optical stellar templates, The Astrophysical Journal Supplement Series 249~(2) (2020) 34.

\bibitem{suwa2018minimum}
Y.~Suwa, T.~Yoshida, M.~Shibata, H.~Umeda, K.~Takahashi, On the minimum mass of neutron stars, Monthly Notices of the Royal Astronomical Society 481~(3) (2018) 3305--3312.

\bibitem{jayasinghe2019asas}
T.~Jayasinghe, K.~Stanek, C.~Kochanek, B.~Shappee, T.~W. Holoien, T.~A. Thompson, J.~Prieto, S.~Dong, M.~Pawlak, O.~Pejcha, et~al., The asas-sn catalogue of variable stars--ii. uniform classification of 412 000 known variables, Monthly Notices of the Royal Astronomical Society 486~(2) (2019) 1907--1943.

\bibitem{mu2022compact}
H.-J. Mu, W.-M. Gu, T.~Yi, L.-L. Zheng, H.~Sou, Z.-R. Bai, H.-T. Zhang, Y.-J. Lei, C.-M. Li, Compact object candidates with k/m-dwarf companions from lamost low-resolution survey, Science China Physics, Mechanics \& Astronomy 65~(2) (2022) 229711.

\bibitem{tucker2024weighing}
M.~Tucker, A.~Wheeler, D.~Rowan, M.~Huber, Weighing the options: The unseen companion in lamost j2354 is likely a massive white dwarf, arXiv preprint arXiv:2407.19004 (2024).

\bibitem{lin2023x}
J.~Lin, C.~Li, W.~Wang, H.~Xu, J.~Jiang, D.~Yang, S.~Yaqup, A.~Iskanda, S.~Ma, H.~Niu, et~al., An x-ray-dim “isolated” neutron star in a binary?, The Astrophysical Journal Letters 944~(1) (2023) L4.

\bibitem{swihart2022new}
S.~J. Swihart, J.~Strader, L.~Chomiuk, E.~Aydi, K.~V. Sokolovsky, P.~S. Ray, M.~Kerr, A new flaring black widow candidate and demographics of black widow millisecond pulsars in the galactic field, The Astrophysical Journal 941~(2) (2022) 199.

\bibitem{conroy2020physics}
K.~E. Conroy, A.~Kochoska, D.~Hey, H.~Pablo, K.~M. Hambleton, D.~Jones, J.~Giammarco, M.~Abdul-Masih, A.~Pr{\v{s}}a, Physics of eclipsing binaries. v. general framework for solving the inverse problem, The Astrophysical Journal Supplement Series 250~(2) (2020) 34.

\bibitem{hobbs2005statistical}
G.~Hobbs, D.~Lorimer, A.~Lyne, M.~Kramer, A statistical study of 233 pulsar proper motions, Monthly Notices of the Royal Astronomical Society 360~(3) (2005) 974--992.

\bibitem{disberg2024deceleration}
P.~Disberg, N.~Gaspari, A.~J. Levan, Deceleration of kicked objects due to the galactic potential, arXiv preprint arXiv:2405.06436 (2024).

\bibitem{bovy2015galpy}
J.~Bovy, galpy: A python library for galactic dynamics, The Astrophysical Journal Supplement Series 216~(2) (2015) 29.

\bibitem{liu2024exploration}
H.~Liu, C.~Du, D.~Ye, J.~Zhang, M.~Deng, Exploration of halo substructures in integrals-of-motion space with gaia data release 3, The Astrophysical Journal 976~(2) (2024) 161.

\bibitem{yan2019chemical}
Y.~Yan, C.~Du, S.~Liu, H.~Li, J.~Shi, Y.~Chen, J.~Ma, Z.~Wu, Chemical and kinematic properties of the galactic disk from the lamost and gaia sample stars, The Astrophysical Journal 880~(1) (2019) 36.

\bibitem{zhu2021element}
H.~Zhu, C.~Du, Y.~Yan, J.~Shi, J.~Ma, H.~J. Newberg, Element abundance analysis of the metal-rich stellar halo and high-velocity thick disk in the galaxy, The Astrophysical Journal 915~(1) (2021) 9.

\bibitem{liao2023hypervelocity}
J.~Liao, C.~Du, H.~Li, J.~Ma, J.~Shi, Hypervelocity stars track back to the galactic center in gaia dr3, The Astrophysical Journal Letters 944~(2) (2023) L39.

\end{thebibliography}

%\begin{thebibliography}{00}

%% For numbered reference style
%% \bibitem{label}
%% Text of bibliographic item

%\bibitem{lamport94}
 % Leslie Lamport,
  %\textit{\LaTeX: a document preparation system},
  %Addison Wesley, Massachusetts,
  %2nd edition,
  %1994.

%\end{thebibliography}
\end{document}